# Limited-angle SPECT image reconstruction using deep image prior

**Kensuke Hori[1], Fumio Hashimoto[2], Kazuya Koyama[1], Takeyuki Hashimoto[3]**

[1] Department of Radiologital Technology, Faculty of Health Science, Juntendo University, 1-5-32, Yushima, Bunkyo-ku, Tokyo, 113-0034, Japan
[2] J. Crayton Pruitt Family Department of Biomedical Engineering, University of Florida, Gainesville, 32611, FL, USA
[3] Department of Radiologital Technology, Faculty of Health Science, Kyorin University, 5-4-1, Shimorenjaku, Mitaka-shi, Tokyo, 181-8612, Japan
*Author to whom any correspondence should be addressed.
E-mail: k.hori.if@juntendo.ac.jp

## Abstract

[Objective] In single-photon emission computed tomography (SPECT) image reconstruction, limited-angle conditions lead to a loss of frequency components, which distort the reconstructed tomographic image along directions corresponding to the non-collected projection angle range. Although conventional iterative image reconstruction methods have been used to improve the reconstructed images in limited-angle conditions, the image quality is still unsuitable for clinical use. We propose a limited-angle SPECT image reconstruction method that uses an end-to-end deep image prior (DIP) framework to improve reconstructed image quality.
[Approach] The proposed limited-angle SPECT image reconstruction is an end-to-end DIP framework which incorporates a forward projection model into the loss function to optimise the neural network. By also incorporating a binary mask that indicates whether each data point in the measured projection data has been collected, the proposed method restores the non-collected projection data and reconstructs a less distorted image.
[Main results] The proposed method was evaluated using 20 numerical phantoms and clinical patient data. In numerical simulations, the proposed method outperformed existing back-projection-based methods in terms of peak signal-to-noise ratio and structural similarity index measure. We analysed the reconstructed tomographic images in the frequency domain using an object-specific modulation transfer function, in simulations and on clinical patient data, to evaluate the response of the reconstruction method to different frequencies of the object. The proposed method significantly improved the response to almost all spatial frequencies, even in the non-collected projection angle range. The results demonstrate that the proposed method reconstructs a less distorted tomographic image.
[Significance] The proposed end-to-end DIP-based reconstruction method restores lost frequency components and mitigates image distortion under limited-angle conditions by incorporating a binary mask into the loss function.

## 1. Introduction

Single-photon emission computed tomography (SPECT) is a nuclear medicine modality that provides metabolic and functional information on the internals of a patient's body [1]. By administering suitable radiopharmaceuticals to patients, SPECT can image various types of information, such as bone metabolism or cerebral blood flow. However, SPECT examinations generally require a long time to obtain sufficient statistical counts because of the limited number of gamma rays emitted by the radiopharmaceuticals. To reduce the burden on the patients, the acquisition time needs to be shortened.

Limited-angle SPECT is a promising acquisition strategy for reducing the number of projections by limiting the projection angle range. Reduction of the projection angle range required for reconstructing a tomographic image allows for a faster scan time and a more flexible scanner design. Faster scan times are particularly beneficial for capturing changes in metabolic or functional activity over time in dynamic scans. Unlike standard SPECT systems that rotate the detector around the target, the detector array in VERITON [2, 3], StarGuide [4], and D-SPECT [5] is arranged in a ring or C-shape, to cover the full-angle range. By reducing the projection angle range from a full-angle to a limited-angle, the number of detector arrays can be decreased in these scanners. The benefits of alleviating the need for a full-angle range include a reduction in the cost of these scanners and a flexible design. Although limited-angle SPECT may produce a paradigm shift in SPECT imaging, challenges remain in image reconstruction.

Limited-angle SPECT data, which lack projections over a certain angle range, causes streak artefacts and image distortions in reconstructed tomographic images, particularly when filtered back-projection (FBP) is used for image reconstruction. Streak artefacts and image distortions result from non-collected projections, which lead to data loss in a bowtie-shaped region of the frequency domain, as posited by the projection-slice theorem. Restoring the lost frequency components is a critical task in suppressing the streak artefacts and image distortion. Iterative reconstruction algorithms such as maximum likelihood expectation maximisation (ML-EM) [6], maximum a posteriori EM (MAP-EM) [7], and a method using reweighted total variation (TV) [8] are known to restore only some of the lost frequency components. Although, iterative image reconstructions can mitigate only some of the streak artefacts and distortions caused by the degradation of spatial resolution along the directions corresponding to the non-collected projection angle range in the tomographic image domain, they still suffer from the artifacts.

Recently, deep learning which achieves superior performance in image reconstruction compared to conventional analytical and iterative reconstruction algorithms [9, 10], has also been used for limited-angle SPECT image reconstruction [11]. Deep learning typically optimises neural network parameters using training datasets in a process known as supervised learning. However, this may not be appropriate for the medical field because of the difficulty of preparing a large number of training datasets for a wide variety of clinical examinations. Deep image prior (DIP) is an alternative approach that trains a neural network using only a target image without requiring large amounts of prior training data [12]. DIP has been widely used in natural image processing and medical imaging owing to its high versatility [13, 14]. In particular, image reconstruction using DIP has attracted attention, with hybrid and end-to-end methods being proposed to address various challenges [15]. In the hybrid reconstruction method, Gong et al. incorporated a DIP framework into conventional penalised image reconstruction using the alternating direction method of multiplier [16], whereby the tomographic image is primarily calculated by penalised iterative image reconstruction and subsequently updated using the DIP framework in the image domain. The advantage of this hybrid method is that it can be used in combination with existing image reconstruction methods. In an end-to-end method, Hashimoto et al. reconstructed tomographic images from measured projection data by incorporating a forward projection model into the DIP framework [17, 18]. An interesting aspect of the end-to-end method is that it does not require a back-projection operation. Instead, a tomographic image is obtained by optimising the neural network parameters through backpropagation. Following this method, Shan et al. reconstructed tomographic images from partial-ring positron emission tomography (PET) data by incorporating the missing projection data regions that are lost because of partial-ring geometry into the end-to-end DIP reconstruction framework [19]. Although end-to-end reconstruction methods have the potential to be applied to both normal cylindrical and partial-ring PET geometry scenarios, their applicability to limited-angle SPECT image has not been explored, highlighting the importance of further investigation to fully exploit their capabilities.

In this study, we propose a limited-angle SPECT image reconstruction method using an end-to-end DIP framework. The proposed method incorporates into the loss function a binary mask representing the non-collected (missing) projection angle regions. Using the binary mask, the proposed DIP reconstruction method functions as an inpainting task in sinogram space, aiding in the recovery of signals in regions excluded by limited-angle acquisition. By analysing the responses in the frequency domain while distinguishing between the collected and non-collected projection angle regions, image distortion suppression performance, that is, the ability to restore the lost frequency components, is clarified. Using both numerical simulations and clinical patient data, the proposed method was compared with conventional iterative and hybrid methods.

## 2. Methodology

### 2.1 SPECT forward projection model

In general, in the SPECT imaging model, the expectation of the projection data $\bar{\boldsymbol{y}} \in \mathbb{R}^N$ is related to the unknown radiotracer activity distribution $\boldsymbol{x} \in \mathbb{R}^M$ through the Radon transform using the system matrix $\boldsymbol{A} \in \mathbb{R}^{N\times M}$:

$$\bar{\boldsymbol{y}} = \boldsymbol{A}\boldsymbol{x}. \tag{1}$$

System matrix $\boldsymbol{A}$ projects from the SPECT image $\boldsymbol{x}$ onto the expected projection data $\bar{\boldsymbol{y}}$. $M$ and $N$ are the number of sampling points in the sinogram and number of voxels in the reconstructed tomographic image, respectively.

### 2.2 DIP

Ulyanov et al. proposed the DIP framework, which enables image restoration tasks without requiring a large real dataset to train the neural network [12]. The training process of DIP denoising uses only a pair of initial data $\boldsymbol{z}$, typically a Gaussian noise image and a degraded measurement image $\boldsymbol{x_0}$, which can be characterised by the following formulation:

$$\boldsymbol{\theta}^* = \min_{\boldsymbol{\theta}} \|f(\boldsymbol{\theta}|\boldsymbol{z}) - \boldsymbol{x_0}\|_2^2, \tag{2}$$

$$\boldsymbol{x}^* = f(\boldsymbol{\theta}^*|\boldsymbol{z}), \tag{3}$$

where $f$ denotes a convolutional neural network (CNN) with trainable parameters $\boldsymbol{\theta}$. Updating $\boldsymbol{\theta}$ to optimal parameters $\boldsymbol{\theta}^*$ using the loss function expressed by equation 2, the DIP framework outputs the denoised image $f(\boldsymbol{\theta}^*|\boldsymbol{z})$ without requiring a noise-free image. To avoid overfitting to the degraded target image, the training process is stopped early.

### 2.3 Proposed method

An overview of the proposed limited-angle SPECT reconstruction method using DIP is shown in Figure 1. The proposed method was implemented in an end-to-end manner, inspired by hybrid DIP PET image reconstruction methods [17, 18]. For end-to-end approach, in equation 2, a tomographic image is defined as $f(\boldsymbol{\theta}|\boldsymbol{z})$, expressing the output from the neural network $f$ with trainable parameters $\boldsymbol{\theta}$ using an initial image $\boldsymbol{z}$. The domain containing the tomographic image is different from the domain containing the measured data which is the measured projection data $\boldsymbol{y}$ in the SPECT system instead of $\boldsymbol{x_0}$. To unify the domain of $f(\boldsymbol{\theta}|\boldsymbol{z})$ and $\boldsymbol{y}$ for the loss function, the SPECT forward projection model $\boldsymbol{A}$ expressed by equation 1 should be implemented as follows:

$$\boldsymbol{\theta}^* = \arg\min_{\boldsymbol{\theta}} \|\boldsymbol{A}f(\boldsymbol{\theta}|\boldsymbol{z}) - \boldsymbol{y}\|_2^2, \tag{4}$$

where $\boldsymbol{A}f(\boldsymbol{\theta}|\boldsymbol{z})$ is the estimated projection using the tomographic image output from the CNN. Limited-angle SPECT cannot acquire a band-shaped portion of the projection data corresponding to regions with limited projection angle range. The CNN estimates the missing data in the measured projection, in short, limited-angle image reconstruction is an inpainting task constrained by equation 1. To aid the recovery of the estimated projection data in non-collected projection angle region, we introduce a binary mask vector $\boldsymbol{l} \in \mathbb{R}^N$ into equation 4 as follows:

$$\boldsymbol{\theta}^* = \arg\min_{\boldsymbol{\theta}} \|\boldsymbol{A}f(\boldsymbol{\theta}|\boldsymbol{z}) \odot \boldsymbol{l} - \boldsymbol{y}\|_2^2, \tag{5}$$

where the operator $\odot$ denotes the Hadamard product. The binary mask $\boldsymbol{l}$ can be preliminarily determined from the scan settings, indicating whether each data region is non-collected (0) or collected (1) in the measured projection data; $\boldsymbol{\theta}^*$ is optimised by solving the minimisation problem in equation 4. The tomographic image $\boldsymbol{x}^*$ is obtained from random noise input $\boldsymbol{z}$, using the neural network in equation 3. Note that the tomographic image is reconstructed from the measured projection data without back-projection operation.

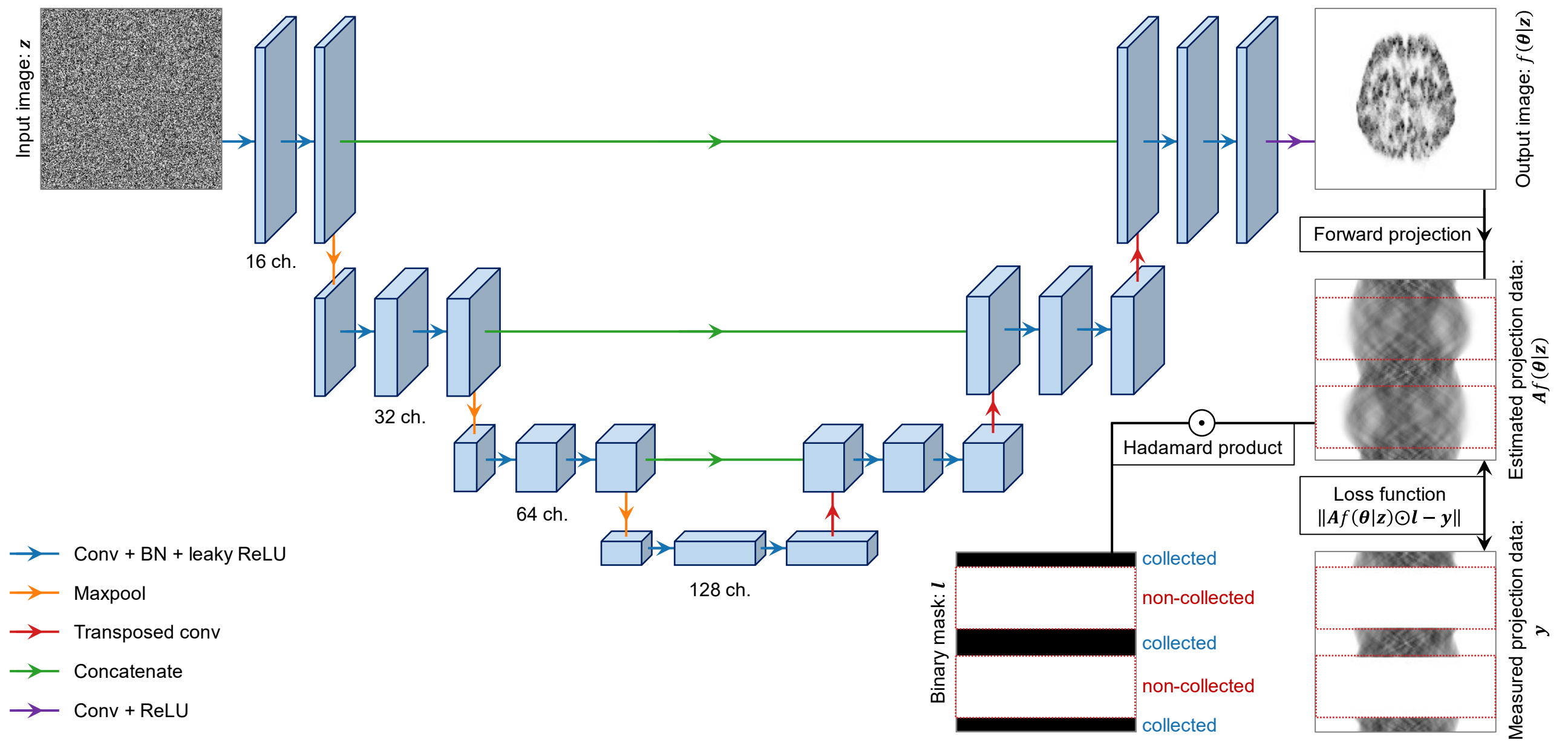

**Figure 1.** Overview of the proposed method. In this CNN model, the random noise image is input and the tomographic image is the output. Each blue box in the neural network corresponds to a feature map with the number of channels shown below. To evaluate the loss function between measured and estimated projection data from the CNN output, the estimated data were binary masked by the missing data region in advance.

### *2.4 Network structure*

In this study, a typical U-net architecture was used for the DIP framework [20]. The CNN consists of an encoder and a decoder (Figure 1). The encoder consists of two components: a 3×3 convolution layer with batch normalisation (BN) and a leaky rectified linear unit (ReLU) activation. The number of feature maps is doubled, and the number of rows and columns is reduced by half for each downsampling layer. The decoder consists of four components: a 4×4 transposed convolution layer with 2 times magnification, concatenation of corresponding feature maps from the encoding part, a 3×3 convolution layer with a BN, and leaky ReLU activation. In this study, the parameters of the CNN were iteratively optimised using Adam; the iteration number was 50,000, and the learning rate was set to $1.0 \times 10^{-4}$. We used Pytorch 1.7.1, and an NVIDIA RTX 2060 graphics processing unit board with 12 GB of memory.

## 3. Experimental setup

### *3.1 Numerical simulations*

We designed 20 numerical phantoms which simulates the concentrations observed in blood flow pattern using BrainWeb phantom [21]. The relative radioactive contrasts of the gray matter, white matter, and cerebrospinal fluid in each slice were assigned as 1, 0.25, and 0.04, respectively [22]. In the numerical simulations, parallel beam projection data were used, assuming the detector response to be a Gaussian function with a full width at half maximum (FWHM) of 10 mm. The attenuation and scatter fractions were ignored. The image size was 128×128×90 voxels with a voxel size of 2.0×2.0×2.0 $\text{mm}^3$. The full projection angle range was 360°, and the limited projection angle range was 120°, specifically 60°-120° and 240°-300°. The projection interval was 4° for both projection angle ranges, and the number of projections was 90.

Two quantitative metrics were employed for image quality evaluation: peak signal-to-noise ratio (PSNR) and structural similarity index measure (SSIM). The PSNR was calculated as follows:

$$\text{PSNR} = 10\log_{10}\left\{\frac{\max(\boldsymbol{\mu})^2}{\|\boldsymbol{\mu} - \widehat{\boldsymbol{\mu}}\|_2^2/N}\right\}, \tag{6}$$

where $\max(\cdot)$ is the maximum value of the image; $\boldsymbol{\mu}$ and $\widehat{\boldsymbol{\mu}}$ are the ground truth and reconstructed tomographic image, respectively; and $N$ is the number of voxels.

Given two image vectors $\boldsymbol{\mu}$ and $\widehat{\boldsymbol{\mu}}$, $K$ kernels $\boldsymbol{k}$ and $\widehat{\boldsymbol{k}}$ with size of 11×11 voxels were sampled, using the following equation for SSIM [23]:

$$\text{SSIM} = \frac{1}{K}\sum_{\boldsymbol{k},\widehat{\boldsymbol{k}}} \frac{(2\mu_{\boldsymbol{k}}\mu_{\widehat{\boldsymbol{k}}} + C_1)(2\sigma_{\boldsymbol{k}\widehat{\boldsymbol{k}}} + C_2)}{(\mu_{\boldsymbol{k}}^2 + \mu_{\widehat{\boldsymbol{k}}}^2 + C_1)(\sigma_{\boldsymbol{k}}^2 + \sigma_{\widehat{\boldsymbol{k}}}^2 + C_2)}, \tag{7}$$

where $\mu_{\boldsymbol{k}}$ and $\mu_{\widehat{\boldsymbol{k}}}$ are the average within each kernel of ground truth and reconstructed tomographic image, respectively; $\sigma_{\boldsymbol{k}}$ and $\sigma_{\widehat{\boldsymbol{k}}}$ are the standard deviations within the kernel of ground truth and reconstructed tomographic image, respectively; $\sigma_{\boldsymbol{k}\widehat{\boldsymbol{k}}}$ is the covariance of $\boldsymbol{k}$ and $\widehat{\boldsymbol{k}}$; $C_1 = (0.01L)^2$ and $C_2 = (0.03L)^2$ are constant values; and $L$ is the dynamic range of the ground truth.

The impact of the spatial frequency of the directions in the collected and non-collected projection angle ranges was evaluated using an object-specific modulation transfer function (O-MTF) [13, 24]. The O-MTF can evaluate how the reconstruction method responds to different frequencies of the object. One-dimensional (1D) O-MTF $m_j$ is represented as follows:

$$m_j = \frac{\sqrt{\sum_{i \in r_j} |\boldsymbol{F\mu}|_i^2}}{\sqrt{\sum_{i \in r_j} |\boldsymbol{F\hat{\mu}}|_i^2}}, \tag{8}$$

where $\boldsymbol{F}$ is the Fourier transform operator; $\boldsymbol{\mu}$ is the reconstructed tomographic image; $\boldsymbol{\hat{\mu}}$ is the reference image; and $\boldsymbol{r}_j$ represents all elements included in the analysis. To calculate the $j$-th element of 1D O-MTF $\boldsymbol{m}$, we first computed the 2D power spectrum for the reconstructed tomographic image $|\boldsymbol{F\mu}|^2$ and reference image $|\boldsymbol{F\hat{\mu}}|^2$. These power spectra were then angularly integrated over $\boldsymbol{r}_j$, which denotes all the elements on a concentric circle of radius $j$. In this study, the properties of spatial frequencies collected in the projection angle ranges of (60°-120° and 240°-300°) or non-collected in ranges of (0°-60°, 120°-240°, and 300°-360°) were evaluated; $\boldsymbol{r}_j$ was also determined by selecting the elements corresponding to these projection angle ranges.

### *3.2 Clinical patient data*

For real SPECT data evaluation, datasets were acquired with 15 min SPECT bone scans on a human subject. The target human was intravenously injected with 740 MBq of $^{99m}$Tc-methylene diphosphonate (MDP). The data were acquired using an Infinia SPECT camera (GE Healthcare) with a set of low-energy, high-resolution parallel-hole collimator. After planar whole-body scintigraphy acquisition, we performed a SPECT scan, in which counts from the 10% energy window at 140 keV were acquired as a 128×128 matrix with a voxel size of 4.42 ×4.42 mm$^2$. The acquisition of a frame of projection data required 10 s and the total number of projections per 360° was 90. Similar to the numerical simulations, the full projection angle range was 360°, and the limited projection angle range was 120°, specifically 60°-120° and 240°-300°. The study was approved by the Research Ethics Committee of Juntendo University (approval number: 23-085).

### *3.3 Comparison algorithms*

The numerical phantoms and clinical patient data were reconstructed using FBP, ML-EM, MAP-EM with TV, Gong's hybrid method [16], and the proposed method. The proposed method was compared with these four methods, to evaluate its performance in restoring the lost frequency components caused by non-collected projections. First, FBP, ML-EM, and MAP-EM with TV are non-deep learning methods. FBP was performed by the use of ramp filter, which has generally been used in clinical practice. For ML-EM and MAP-EM with TV, the number of iterations was 100. In particular, MAP-EM with TV is often employed for image reconstruction from insufficient projection data to suppress image noise or artefacts while preserving the edges. We used Green's one-step late algorithm for MAP-EM, and the regularisation parameter was set to $1.0\times10^{-3}$.

The hybrid and proposed methods are both unsupervised deep learning methods that use the DIP framework. In the hybrid method, the measured projection data are first transformed into tomographic images using penalised image reconstruction, and then the DIP framework is applied to the image domain rather than the projection domain. The domain in which the DIP framework is applied is different from that of the proposed method, which was developed as an end-to-end approach.

Regarding the technical difference in the transformation of projection data into tomographic images, the comparison algorithms utilise back-projection, whereas the proposed method employs the DIP framework. Here, we investigated the performance of the DIP framework applied in image domain and projection domain.

## 4. Results

### *4.1 Numerical simulations*

Figure 2 shows three orthogonal slices of the reconstructed results from the simulation dataset obtained using different reconstruction algorithms. Compared with the other methods, the proposed method reconstructed the tomographic images without streak artefacts and image distortions. Furthermore, the gray matter shape reconstructed by the proposed method is closer to the ground truth than that reconstructed by the other methods. Figure 3 shows the box plots of PSNR and SSIM for the different reconstruction algorithms. The average quantitative values for the FBP, ML-EM, MAP-EM with TV, the hybrid, and proposed method are 19.81, 32.42, 32.44, 32.15, and 33.28 dB, respectively, in terms of PSNR, and 0.7794, 0.9441, 0.9442, 0.9438, 0.9527, respectively, in terms of SSIM. Both the PSNR and SSIM of the proposed method are significantly higher than those of the other methods. Figure 4 shows the power spectra of the reconstructed axial images displayed in Figure 2 for the different reconstruction algorithms. The frequency components are lost in the non-collected projection angle region, which is the region outside the bowtie indicated by the green lines in the power spectrum, for the ML-EM, MAP-EM with TV, and hybrid methods. Whereas the lost frequency components in the non-collected projection angle region are restored in the tomographic image reconstructed by the proposed method. The O-MTFs in Figure 5 evaluate how the reconstruction method responds to different frequencies of the object in the collected (60°-120° and 240°-300°) and non-collected projection angle ranges (0°-60°, 120°-240°, and 300°-360°). The O-MTF curves for the different reconstruction algorithms exhibit almost no change in the collected projection angle range. O-MTF represents the ratio of the square root of the angular-averaged 2D power spectra of the original and reconstructed tomographic images, with a value closer to 1 indicating better performance. Considering this specification, in the non-collected projection angle range, the O-MTF curve for the proposed method is remarkably improved compared with those of the other methods.

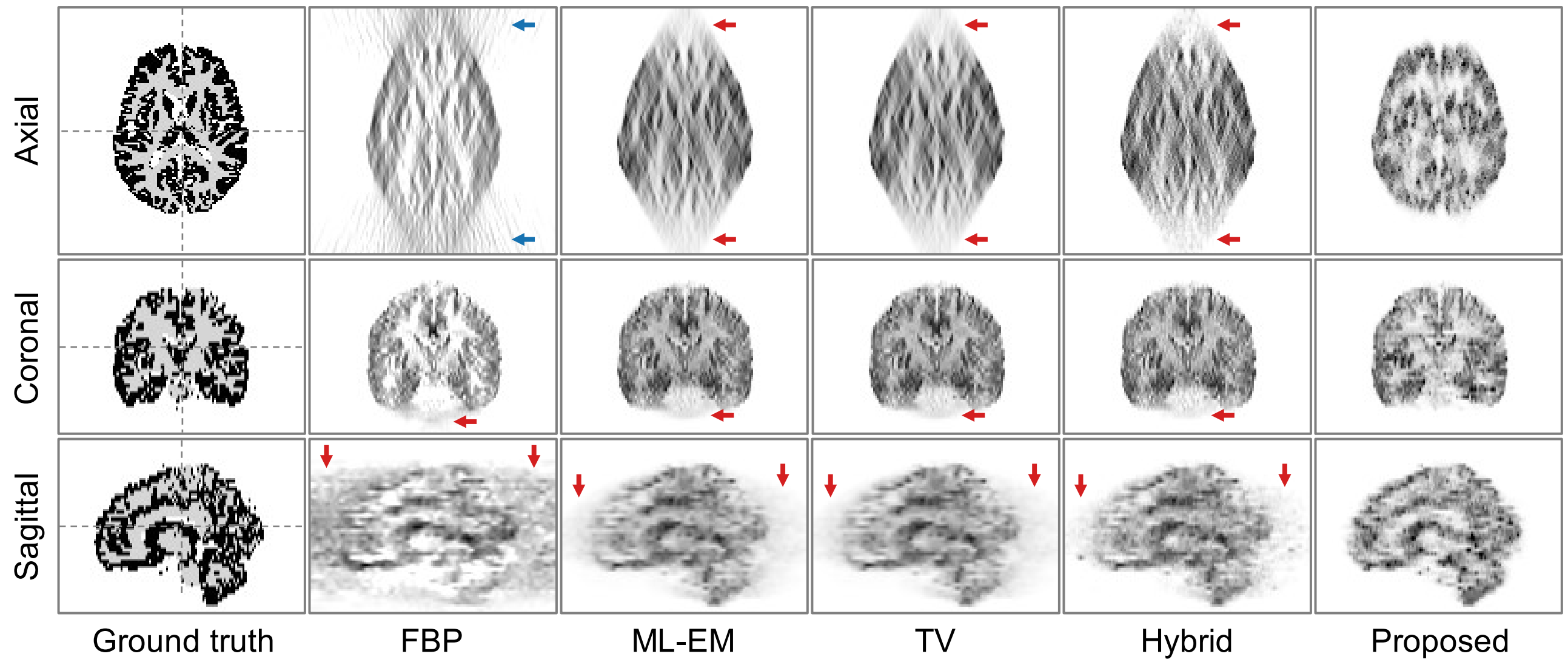

**Figure 2.** The reconstructed tomographic images for the BrainWeb data for different reconstruction algorithms; FBP, ML-EM, MAP-EM with TV, hybrid, and proposed methods. The limited-angle condition was achieved by collecting projection data only from the 60°-120° and 240°-300° range. The blue and red arrows indicate streak artefacts and image distortions, respectively.

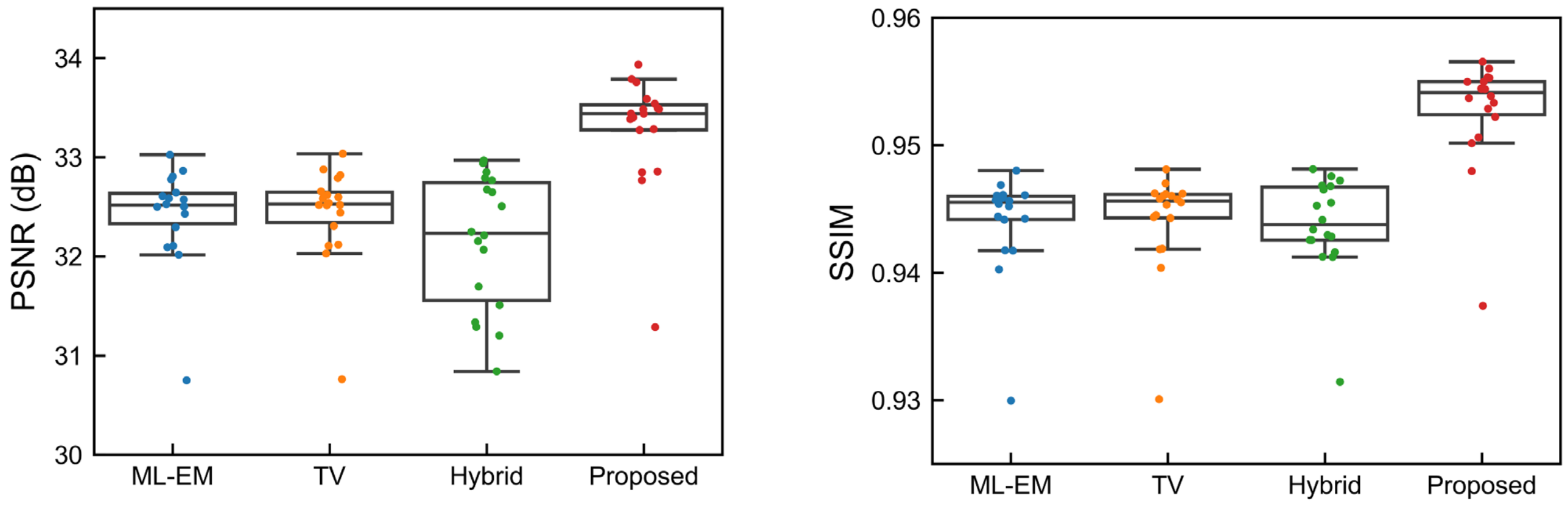

**Figure 3.** Box plots of the reconstructed tomographic images in terms of PSNR (left) and SSIM (right) using different reconstruction algorithms. The horizontal line in each box represents the median value. The lower and upper edges of each box correspond to the first and third quartiles, respectively, and the lower and upper whiskers indicate the minimum and maximum values within 1.5 times the interquartile range from these quartiles, respectively.

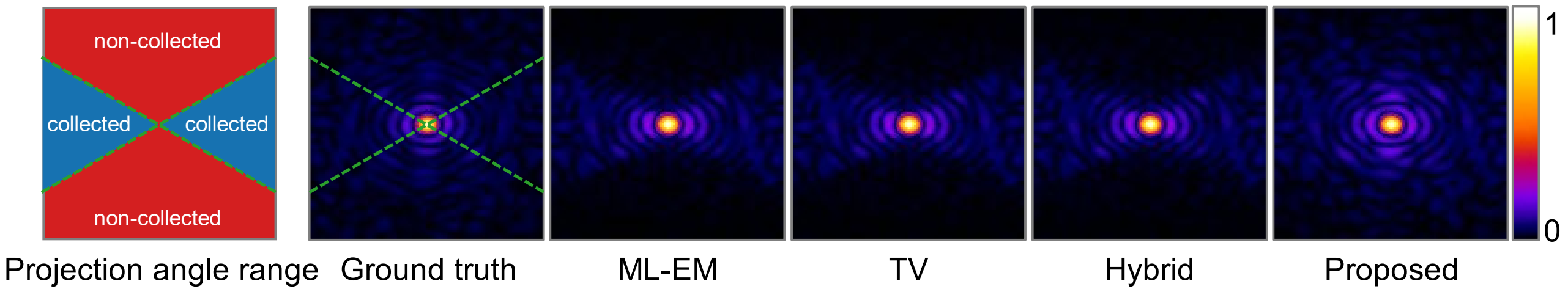

**Figure 4.** Power spectrum of the reconstructed BrainWeb data using different reconstruction algorithms. The leftmost image shows the frequency space in which projection data were either collected or omitted under the limited-angle condition, collecting projection data within the 60°-120° and 240°-300° range. Green lines in leftmost image and ground truth indicate the boundary between collected and non-collected projection angle range.

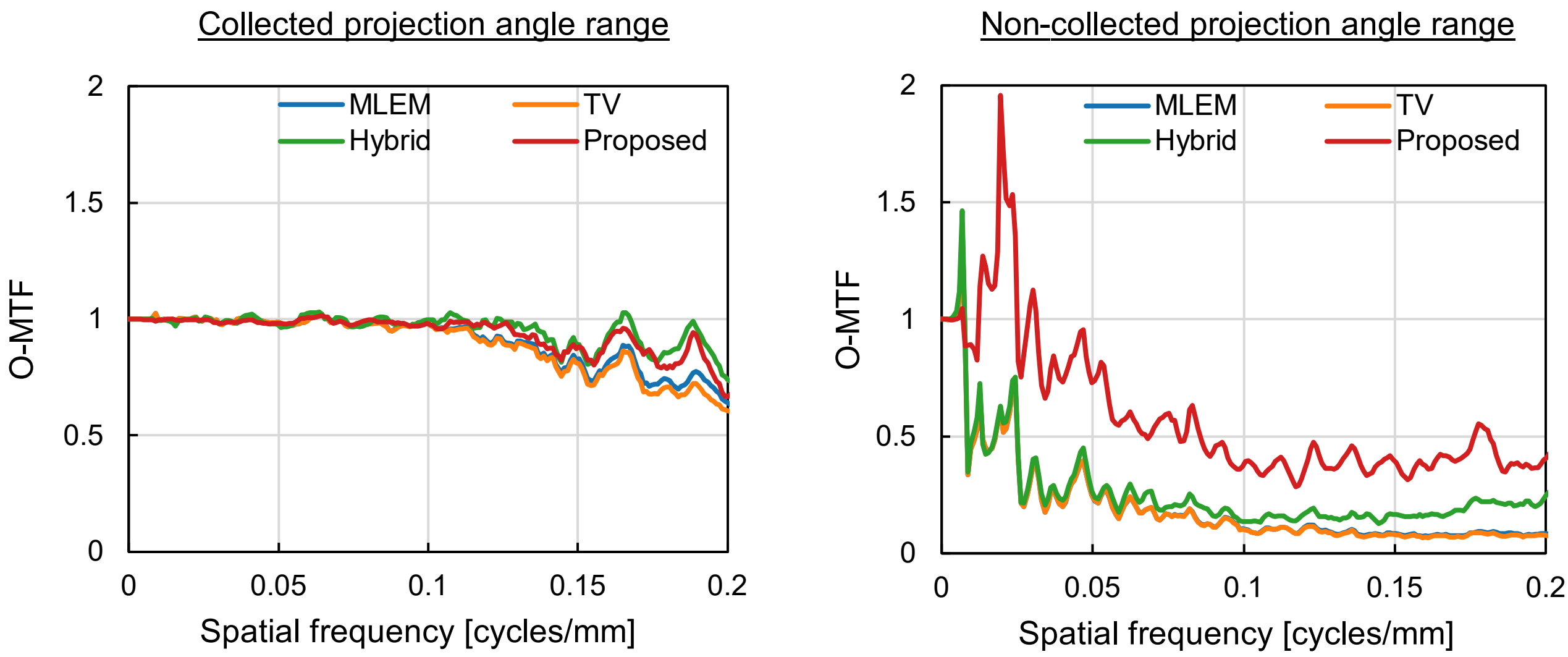

**Figure 5.** O-MTFs of the reconstructed BrainWeb data using different reconstruction algorithms. Left and right are the results evaluated for the collected (60°-120° or 240°-300°) and non-collected projection angle ranges (0°-60°, 120°-240°, or 300°-360°), respectively.

### *4.2 Clinical patient data*

Figure 6 shows three orthogonal slices of the reconstructed results from the clinical patient data obtained using different reconstruction algorithms. In the axial images reconstructed using FBP, ML-EM, MAP-EM with TV, and the hybrid method, the sternal bone, indicated by a red arrow, is distorted in the anterior-posterior direction. Compared with existing image reconstruction methods, the tomographic images reconstructed by the proposed method are distortion free. Figure 7 shows the line profiles, obtained using different reconstruction algorithms, in the anterior-posterior direction through the sternal bone. The FWHMs for ML-EM, MAP-EM with TV, hybrid, and proposed methods are 23.21, 38.0, 38.07, 33.57, and 26.71 mm, respectively. Figure 8 shows the power spectra for the reconstructed tomographic images in Figure 6 using different reconstruction algorithms. As with the results of the numerical simulations, compared with the other methods, the proposed method restored the frequency components even in the non-collected projection region which is indicated by the red region. Figure 9 shows the O-MTFs of the reconstructed tomographic images, which are the axial images shown in Figure 6. The O-MTF curves for the different reconstruction algorithms show almost no change in the collected projection angle range (60°-120° and 240°-300°). In the non-collected projection angle range (0°-60°, 120°-240°, and 300°-360°), the O-MTF curve for the proposed method is improved compared with that of the other methods.

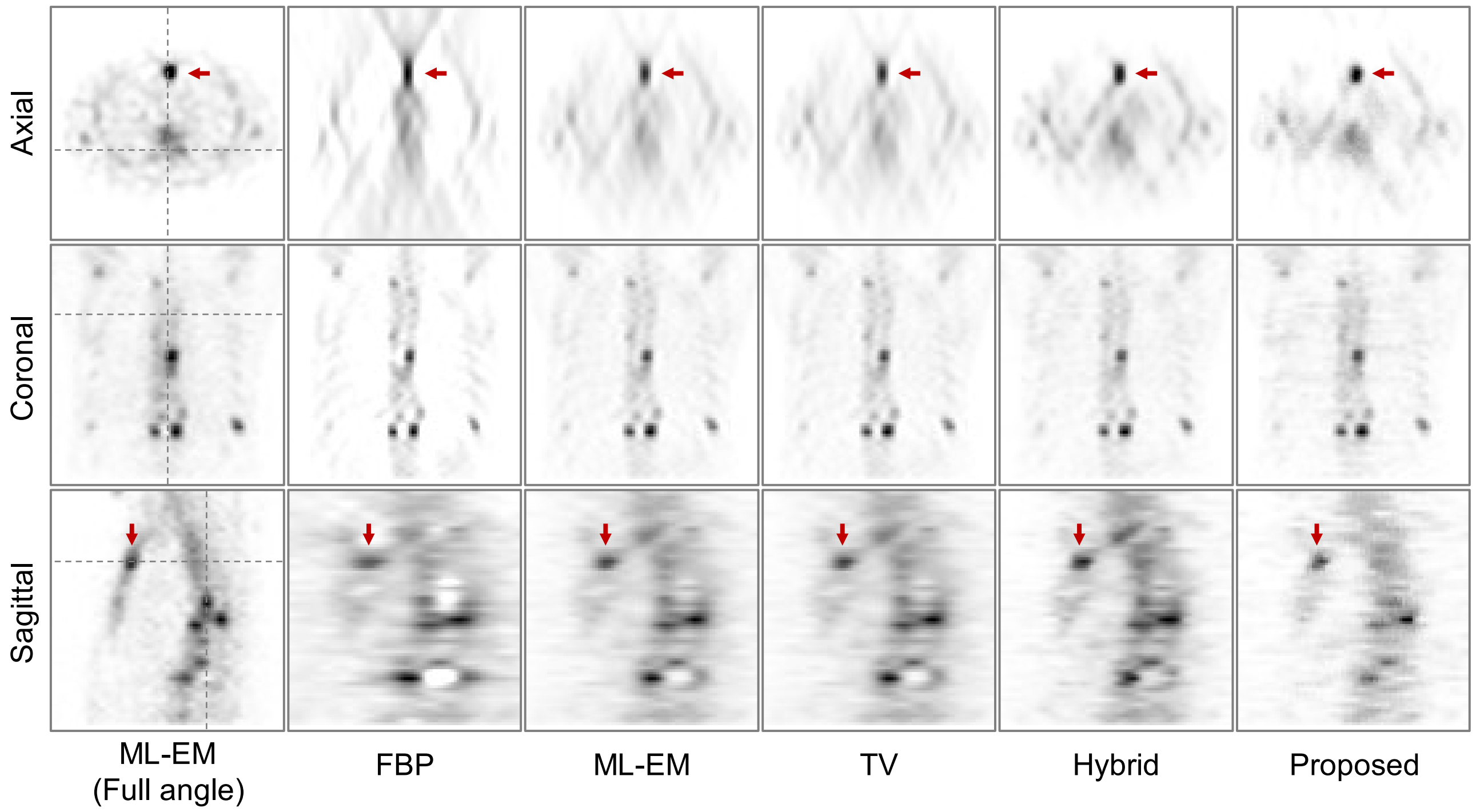

**Figure 6.** Reconstructed images for the clinical patient data for different reconstruction algorithms; FBP, ML-EM, MAP-EM with TV, the hybrid, and proposed methods. The limited-angle condition was achieved by collecting projection data in the 60°-120° and 240°-300° range. The red arrows indicate image distortion in the sternal bone.

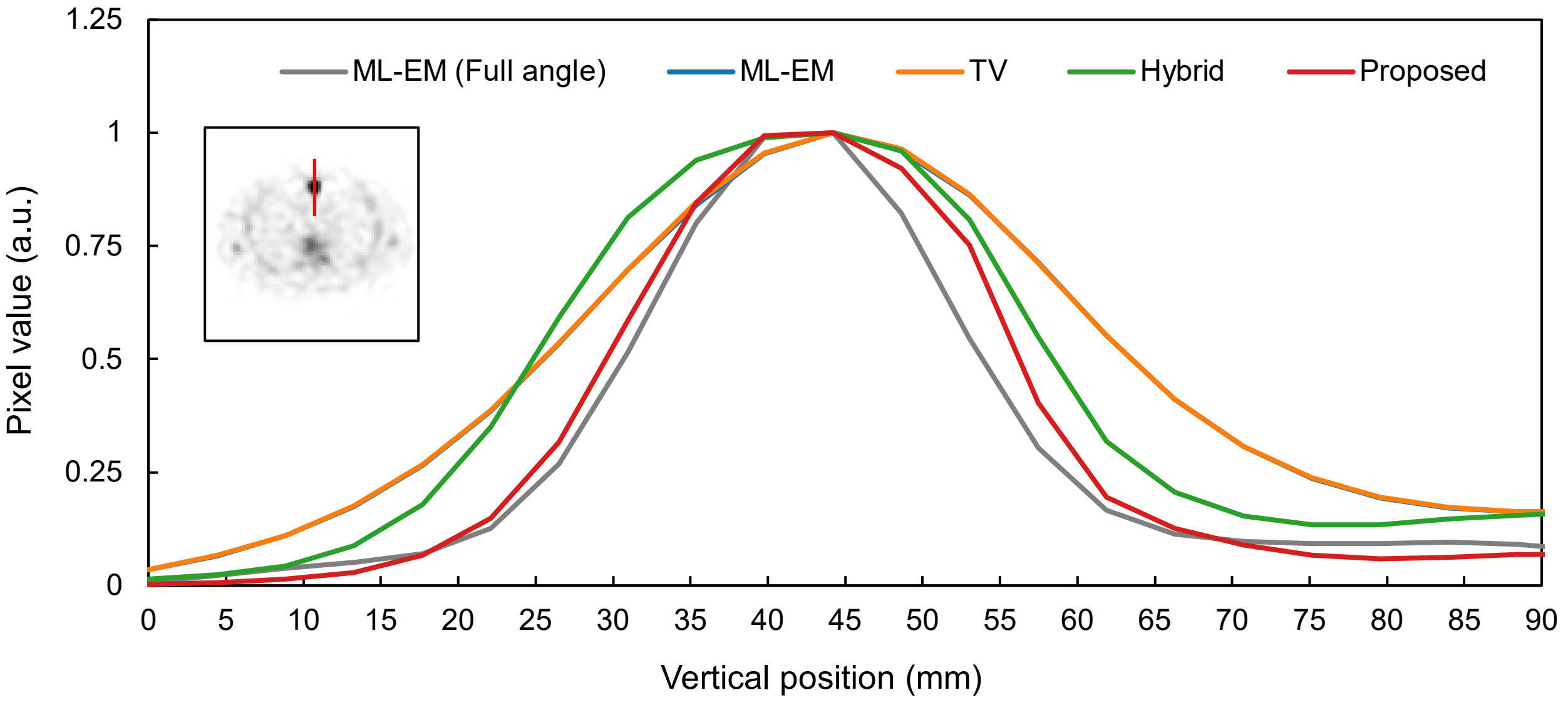

**Figure 7.** Line profiles of the reconstructed clinical patient data are shown in the top row of Figure 6 for different reconstruction algorithms; ML-EM (full angle), MAP-EM with TV, hybrid, and proposed methods. The profile line is shown in the upper left of the graph.

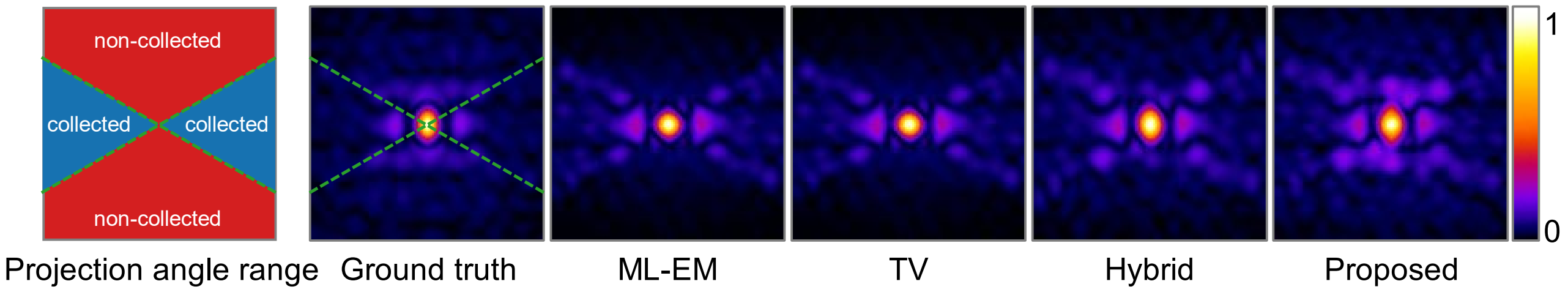

**Figure 8.** Power spectrum of the reconstructed clinical patient data using different reconstruction algorithms. The leftmost image shows the frequency space in which projection data were either collected or omitted under the limited-angle condition, collecting projection data within the 60°-120° and 240°-300° range. Green lines in the leftmost image and ground truth indicate the boundary between collected and non-collected projection angle range.

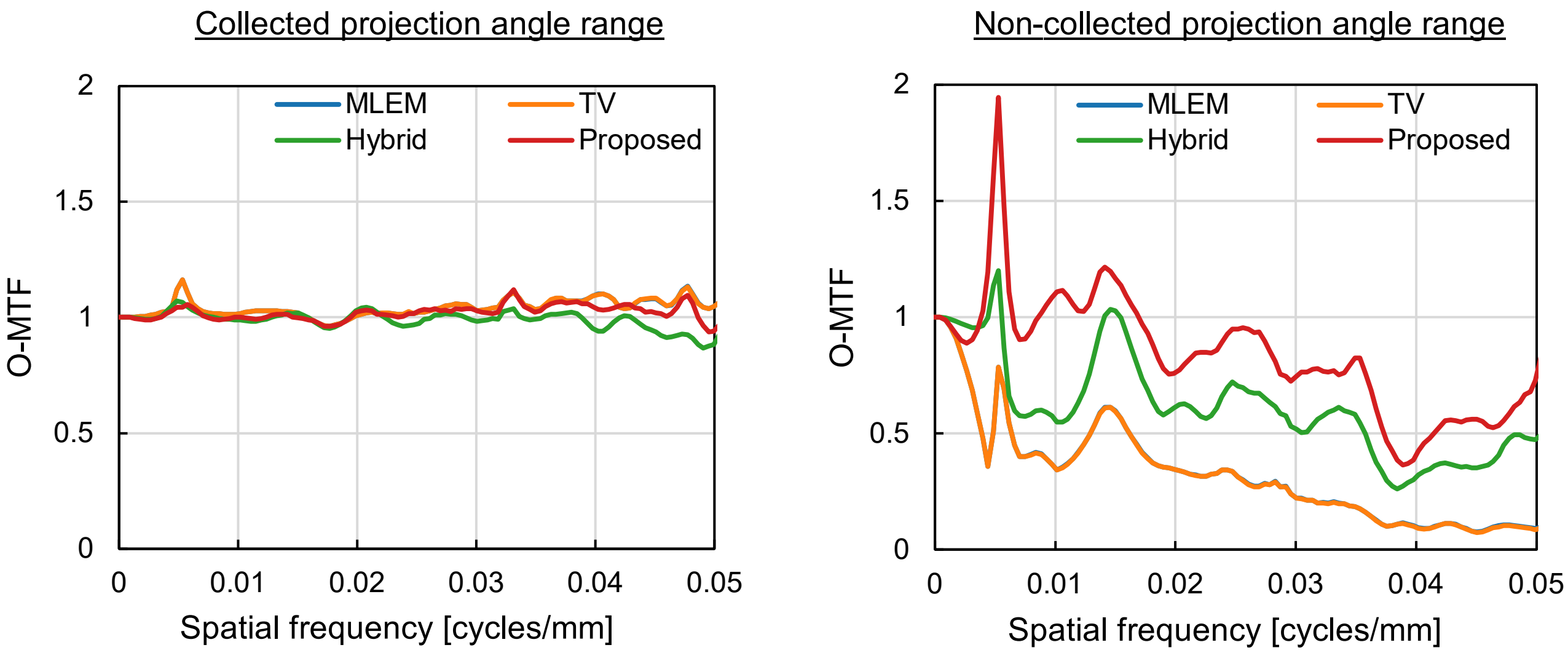

**Figure 9.** O-MTFs of the reconstructed the clinical patient data with different reconstruction algorithms. Left and right are the results evaluated for collected (60°-120° or 240°-300°) and non-collected projection angle range (0°-60°, 120°-240°, or 300°-360°), respectively. The O-MTFs for MLEM are almost the same as those for MAP-EM with TV.

## 5. Discussions

In the existing image reconstruction methods, the transformation from the projection to the image domain is achieved by back-projection as a physically inverse process of projection data measurement. The back-projection at a certain angle in the image domain and filling of the appropriate values in pixels on the same angle line in the frequency domain are equivalent according to the projection-slice theorem. In the limited-angle condition, the non-collected projection angles make it difficult to fill or update the pixel values along the corresponding directions in the frequency domain because there are no projection data to be backprojected in the image domain. Existing image reconstruction methods using back-projection have difficulty restoring lost frequency components, as posited by the projection-slice theorem. In contrast, the proposed method uses backpropagation instead of back-projection to update the neural network parameters which indirectly represent the reconstructed tomographic image. By updating the neural network parameters with backpropagation which is not based on the projection-slice theorem, the lost frequency components were remarkably restored using the proposed method compared with existing image reconstruction methods, as shown in Figures 4, 5, 8, and 9. Comparing the proposed method with the hybrid method clarifies the reason for the remarkable frequency restoration; the technical difference between the proposed and hybrid method is using backpropagation versus back-projection for transformation from the projection domain to the image domain. Although the hybrid method also implements the DIP framework, the output image of the neural network converges to a distorted image that has already been reconstructed by employing a penalised image reconstruction. These findings indicate that frequency restoration by the neural network implemented in an end-to-end approach yields a less distorted tomographic image than the image reconstructed using back-projection because it does not rely on the projection-slice theorem. In other words, we found that end-to-end neural network optimisation can restore the lost frequency components better than back-projection under limited-angle conditions because of an inpainting task in the sinogram space leading to the recovery of non-collected projection data.

We performed not only the numerical simulations in ideal situation but also clinical patient data in real conditions including complex physical phenomena such as scattering and attenuation. The proposed method may be effective in clinical practice because the lost frequency components can be restored more precisely than using existing image reconstructions in both experiments. In these experiments, it was observed that less distorted tomographic images can be reconstructed even if the projection angle range is reduced by a third (from 360° to 120°). By reducing the projection angle range in standard SPECT systems, in which the detectors rotate around the target to acquire the projection data, the scan time is reduced and the risk of examination interruption caused by unexpected patient movement is mitigated. Focusing on specialised SPECT systems [2-5], such as VERITON, StarGuide, and D-SPECT, in which detectors do not rotate around the target, the proposed method may reduce the number of detector arrays by alleviating the need for projection data in the full-angle range. The fewer the number of detector arrays implemented, the more cost-effective and flexible is the SPECT system. In another example, multi-pinhole SPECT gets into limited-angle condition. Even though multi-pinhole SPECT simultaneously acquires multiple projections, the projection angle limitation forces the detector to rotate, as in standard SPECT systems. The proposed method has the potential to eliminate the need for rotating the detector in a multi-pinhole SPECT system.

When the learning rate is increased to reduce the calculation time, convergence to the optimal solution is difficult with the proposed method, which uses an end-to-end DIP framework for limited-angle image reconstruction. To achieve stable convergence to the optimal solution, in this study, the learning rate was set to a very small value ($1.0\times10^{-4}$) which resulted in an excessively long computation time for clinical applications. By applying a block iterative algorithm [18] or transferring the pre-trained neural network parameters to the end-to-end DIP framework [25-27], the calculation time of the proposed method can be made suitable for clinical applications even for a low learning rate. The block iterative algorithm not only increases the computation time of the proposed method but also reduces the GPU memory usage [18]. Although the tomographic image was reconstructed slice-by-slice in this study, by reducing GPU memory usage, the proposed method may be used to perform a fully 3D reconstruction which is required for multi-pinhole SPECT. By transferring the neural network parameters which are pre-trained using the tomographic image reconstructed by the back-projection-based method, the calculation time of the proposed method can be further reduced [25]. However, the back-projection-based method used in pre-training may encourage the solution to converge to a distorted tomographic image. To address this concern, during pre-training, data other than the target data can be reconstructed using the proposed method instead of back-projection [26, 27].

In the clinical practice of nuclear medicine, imaging diagnosis is performed by quantitative analysis of radioisotopes accumulated in organs or lesions. Specifically, the amount of accumulated radioisotopes, estimated from image pixel values, is used for the quantitative analysis of metabolic and functional information. Without a linear relationship between radioactivity concentration and image pixel value, the results of the quantitative analysis may lead to misdiagnosis. By evaluating the linearity of the proposed method, its potential for clinical application can be enhanced. As an example of a future clinical application, bone SPECT which is added to 2D bone scintigraphy, is mentioned in this study. Based on a 2D bone scintigraphy scan of 10-15 minutes, image diagnosis tends to be difficult because of the superimposed radioactivity concentrations. While bone SPECT reconstructs the tomographic image which can improve a diagnostic accuracy compared with the 2D bone scintigraphy, the scan time for a whole body is too long (30-45 minutes) to be a routine work. The proposed method reduces the scan time of bone SPECT. Consequently, the tomographic image can be obtained routinely, improving bone scan accuracy. In the proposed method, the neural network was optimised using the target data, and its application to other body parts should be clarified through further clinical studies.

## 6. Conclusion

In this study, we proposed a limited-angle image reconstruction method using an end-to-end DIP framework. The proposed method introduces a forward projection model and a binary mask, representing the non-collected projection angle range, to a loss function, to achieve innovative image reconstruction without back-projection. Compared with FBP, ML-EM, MAP-EM with TV, and hybrid methods, the proposed method can suppress image distortions in both numerical simulations and clinical patient data. Furthermore, by analysing not only in image domain but also in frequency domain, the results show that the proposed method outperforms the back-projection-based method in terms of frequency component restoration. Future work can expand the application of the proposed method to include other organs and imaging purposes.

## Acknowledgements

The authors would like to thank Mr. Tomoya Harada from the LSI Sapporo clinic for acquiring the clinical patient data.

## Data availability statement

Data cannot be made publicly available upon publication because they are not available in a format that is sufficiently accessible or reusable by other researchers. The data supporting the findings of this study are available upon reasonable request from the authors.

## ORCID iDs

Kensuke Hori: https://orcid.org/0009-0001-8269-7272
Fumio Hashimoto: https://orcid.org/0000-0003-2352-0538
Kazuya Koyama: https://orcid.org/0000-0002-8060-9793
Takeyuki Hashimoto: https://orcid.org/0000-0001-6465-0836